\documentclass[%
 aps, pra, reprint,
 superscriptaddress, 
 showpacs, nopreprintnumbers,
 longbibliography,
 amsmath, amssymb, floatfix
]{revtex4-1}

\usepackage{graphicx}
\usepackage{dcolumn}
\usepackage{bm}
\usepackage{hyperref}
\usepackage{physics}
\usepackage[usenames,dvipsnames]{xcolor}

\newcommand{\stateup}{\ensuremath{\uparrow}}
\newcommand{\statedn}{\ensuremath{\downarrow}}

\newcommand{\z}{\ensuremath{0}}
\newcommand{\one}{\ensuremath{1}}
\newcommand{\zz}{\ensuremath{00}}
\newcommand{\zo}{\ensuremath{01}}
\newcommand{\oz}{\ensuremath{10}}
\newcommand{\oo}{\ensuremath{11}}

\newcommand{\bmag}{\ensuremath{B}}
\newcommand{\Bres}{\ensuremath{B_0}}
\newcommand{\Brot}{\ensuremath{B_{\textrm{rot}}}}

\newcommand{\Eave}{\ensuremath{\overline{E}}}
\newcommand{\edmtr}{\ensuremath{\bm{d}_{10}}}
\newcommand{\edm}{\ensuremath{\bm{d}}}
\newcommand{\edmtrabs}{\ensuremath{{d}_{10}}}

\newcommand{\ftrap}{\ensuremath{f_\mathrm{trap}}}
\newcommand{\Hmol}{\ensuremath{H_{\text{mol}}}}

\newcommand{\Hmw}{\ensuremath{H_{\text{mw}}}}

\newcommand{\Hddi}{\ensuremath{H_{\text{ddi}}}}
\newcommand{\Htwomol}{\ensuremath{H_{\text{2mol}}}}
\newcommand{\Imat}{\ensuremath{{\cal I}}}
\newcommand{\Itrap}{\ensuremath{I_\mathrm{tr}}}
\newcommand{\Vddi}{\ensuremath{V_{\text{ddi}}}}
\newcommand{\omol}{\ensuremath{\omega_{\text{mol}}}}
\newcommand{\omw}{\ensuremath{\omega_{\text{mw}}}}
\newcommand{\Ugate}{\ensuremath{\mathcal{T}}}
\newcommand{\Udesired}{\ensuremath{U_\mathrm{ideal}}}

\newcommand{\Umot}{\ensuremath{U_\mathrm{motion}}}
\newcommand{\tgate}{\ensuremath{\tau_\mathrm{gate}}}
\newcommand{\trms}{\ensuremath{\tau_\mathrm{rms}}}

\newcommand{\mytitle}{A robust entangling gate for polar molecules using magnetic and microwave fields}

\begin{document}

\title{\mytitle}

 \author{Michael Hughes}
  \email{michael.hughes@physics.ox.ac.uk}
  \affiliation{Clarendon Laboratory, University of Oxford, Parks Rd, Oxford OX1 3PU, United Kingdom.}

 \author{Matthew D. Frye} 
 \affiliation{Joint Quantum Centre (JQC) Durham-Newcastle, Department of Chemistry, Durham University, South Road, Durham DH1 3LE, United Kingdom}
 
 \author{Rahul Sawant}
 \affiliation{Joint Quantum Centre (JQC) Durham-Newcastle, Department of Physics, Durham University, South Road, Durham DH1 3LE, United Kingdom}

 \author{Gaurav Bhole}
 \affiliation{Clarendon Laboratory, University of Oxford, Parks Rd, Oxford OX1 3PU, United Kingdom.}

 \author{Jonathan A. Jones}
 \affiliation{Clarendon Laboratory, University of Oxford, Parks Rd, Oxford OX1 3PU, United Kingdom.}

 \author{Simon L. Cornish}
 \affiliation{Joint Quantum Centre (JQC) Durham-Newcastle, Department of Physics, Durham University, South Road, Durham DH1 3LE, United Kingdom}

 \author{M. R. Tarbutt}
 \affiliation{Centre for Cold Matter, Blackett Laboratory, Imperial College London, Prince Consort Road, London SW7 2AZ, United Kingdom.}

 \author{Jeremy M. Hutson}
 \affiliation{Joint Quantum Centre (JQC) Durham-Newcastle, Department of Chemistry, Durham University, South Road, Durham DH1 3LE, United Kingdom}
 
 \author{Dieter Jaksch} 
 \affiliation{Clarendon Laboratory, University of Oxford, Parks Rd, Oxford OX1 3PU, United Kingdom.}
 
 \author{Jordi Mur-Petit} 
 \email{jordi.murpetit@physics.ox.ac.uk}
 \affiliation{Clarendon Laboratory, University of Oxford, Parks Rd, Oxford OX1 3PU, United Kingdom.}

\date{\today}

\begin{abstract}
 Polar molecules are an emerging platform for quantum technologies based on their long-range electric dipole--dipole interactions, which open new possibilities for quantum information processing and the quantum simulation of strongly correlated systems.
 Here, we use magnetic and microwave fields to design a fast entangling gate with $>0.999$ fidelity and which is robust with respect to fluctuations in the trapping and control fields and to small thermal excitations.
 These results establish the feasibility to build a scalable quantum processor with a broad range of molecular species in optical-lattice and optical-tweezers setups.
\end{abstract}

\maketitle


\section{Introduction}

The field of ultracold molecules has seen enormous progress in the last few years, with landmark achievements such as
the production of the first quantum-degenerate molecular Fermi gas~\cite{DeMarco2019},
low-entropy molecular samples in optical lattices~\cite{Moses2015,Reichsollner2017},
trapping of single molecules in optical tweezers~\cite{Liu2018,Anderegg2019,Liu2019},
and magneto-optical trapping and sub-Doppler cooling of molecules~\cite{Barry2014, Truppe2017, Williams2018, Anderegg2018, Collopy2018}.
These results bring significantly closer a broad range of applications of ultracold molecules, from state-controlled chemistry~\cite{Balakrishnan2001, Tscherbul2006jcp, Ospelkaus2010, Wolf2017, Gregory2019, Hu2019} and novel tests of fundamental laws of nature~\cite{Flambaum2007,DeMille2008, Baron2014, Safronova2018}
to new architectures for quantum computation~\cite{DeMille2002, Yelin2006, Mur-Petit2013atmol, Sawant2020, Albert2019}, quantum simulation~\cite{Micheli2006, Gorshkov2011, Gorshkov2013, Covey2018, Blackmore2019, Rosson2020} and quantum sensing~\cite{Alyabyshev2012,Mur-Petit2015}.

A key feature of polar molecules 
is the strong long-range electric dipole--dipole interaction (DDI) between them.
Full exploitation of this feature requires tools to tune the DDI, by controlling the underlying molecular electric dipole moments (EDMs).
A popular approach to this involves trapping the molecules in a two-dimensional array, which could be an optical lattice~\cite{Ospelkaus2010, Yan2013} or an array of optical tweezers~\cite{Liu2018, Anderegg2019, Liu2019}.
A static electric field mixes the rotational states~\cite{Friedrich1991,Friedrich1991nat}, leading to an EDM dependent on the external field.
The field needed to produce a laboratory-frame EDM close to the molecule-frame EDM, $d$, is $E_\text{app} \simeq \Brot/d$. For heavy bialkali-metal molecules, whose rotational constants, $\Brot$, are small, 
$E_\mathrm{app}\approx 1$~kV/cm, which is easy to achieve.
For other molecules, especially hydrides, the required field can be hundreds of times larger, which is challenging.
Another limitation of this approach is that the induced EDM depends on
the strength of the polarising field, making the DDI between molecules highly sensitive to errors or fluctuations in this field.

Here, we describe an alternative approach to controlling the electric DDI that does not involve static electric fields, but relies instead on magnetic and microwave (MW) fields~\cite{Gorshkov2011, Pellegrini2011, Herrera2014, Ni2018}.
We employ this tunable DDI together with a shaped MW pulse~\cite{Benhelm2008nphys, Schafer2018, Barends2014, Li2019} to design an entangling two-qubit gate that has a large fidelity, $\mathcal{F}>0.999$, and is robust with respect to experimental uncertainties in the optical confinement and to thermal motional excitations.
Given that our achievable fidelity is above the quantum-error-correction threshold~\cite{Knill2005, Reichardt2006}, these results establish the feasibility of universal fault-tolerant quantum computation~\cite{Brylinski2002, Bremner2002} with a wide range of polar molecules in scalable platforms.
For concreteness, we illustrate our discussion with numerical results for CaF (X$^2\Sigma$), which has been laser-cooled to temperatures below 10~$\mu$K~\cite{Truppe2017, Williams2018, Anderegg2018,
Cheuk2018, Anderegg2019, Blackmore2019}.
Our proposal is also applicable to bialkali-metal molecules in their lowest $^1\Sigma$ or $^3\Sigma$ states;
to illustrate this, we present in Appendix~\ref{sec:RbCs} analogous numerical results for RbCs~\cite{Takekoshi2014, Molony2014, Gregory2016, Gregory2017, Reichsollner2017, Gregory2019}.

\section{Controlling the molecular EDM with magnetic and MW fields}

\begin{figure}[tb] 
	\centering
	\includegraphics[width=\columnwidth]{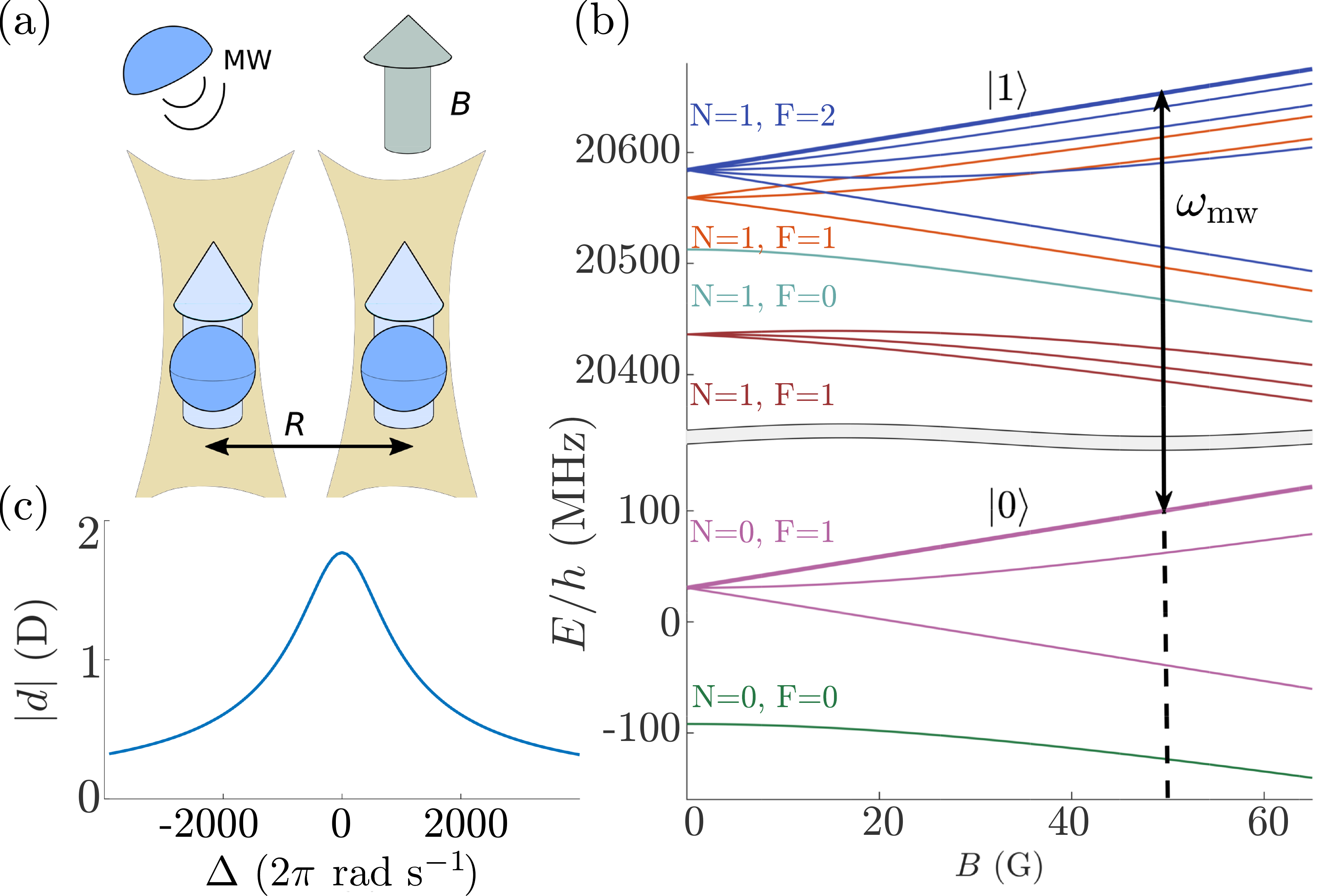}
	\caption{ 
	(a) Sketch of the system under consideration: two polar molecules (spheres with arrows) trapped in optical tweezers (shaded areas) and subject to magnetic, $\bmag$, and microwave (MW) fields.
	(b) Energy levels
	of the $N=0$ and $N=1$ rotational manifolds of the X$^2\Sigma^+(v=0)$ state of $^{40}$Ca$^{19}$F as a function of magnetic field.
	The arrow indicates the transition addressed by the MW field to dress the states
	$\ket{\z} = \ket{N=0, F=1, M_F=1}$
	and 
	$\ket{\one} = \ket{N=1, F=2, M_F=2}$,
	where $F$ is the total angular momentum quantum number of the molecule, and $M_F$ its projection on the $z$ axis, defined by the magnetic field.
	At $\Bres=50$~G, the resonant transition frequency, $\omol$, is 20.778~GHz and the transition EDM is $\edmtrabs\approx 1.77$~D.
    (c) Absolute value of the EDM of the eigenstates of $\Hmol$
    as a function of detuning for a constant Rabi frequency $\Omega = 2\pi\times 731$~rad/s.}
	\label{fig:CaF-levels}
\end{figure}

The first step in processing quantum information
with polar molecules is to isolate a pair of levels to define a qubit space. 
To this end, we apply a homogeneous magnetic field of magnitude $\Bres$ to separate the Zeeman components of the fine and hyperfine levels within a rotational manifold, and MWs to couple a selected Zeeman state to a state in an adjacent rotational manifold~\cite{Gorshkov2011}.
We show in Fig.~\ref{fig:CaF-levels} the energies of the states in the $N=0$ and 1 rotational manifolds of CaF in a magnetic field $\bmag$, with $N$ the rotational quantum number.
For $\bmag >30$~G, the different Zeeman states within a rotational manifold are split by $\gtrsim 10$~MHz.
This large splitting allows selected states within $N=0$ and $N=1$ to be coupled using MW radiation with negligible off-resonant excitation to other states, thus defining a qubit space, $\{ \ket{\one},\ket{\z} \}$.
In the absence of a static electric field,
the qubit states
satisfy $\braket{j|\edm}{j}=0$ $(j=\z,\one)$, while $\braket{\one|\edm}{\z}=\edmtr$, the transition EDM. 
They can be resonantly coupled by suitably polarised MWs of angular frequency $\omw \approx \omol$, where $\hbar\omol(\Bres)=E_1(\Bres)-E_0(\Bres)$,
with $E_j(B)$ the energy of state $\ket{j}$ as a function of $B$. 
We also introduce $\Eave(B) = (E_0(B) + E_1(B))/2$.
In the electric dipole approximation, the MW coupling is
$\Hmw = -\edm\cdot\bm{E}$
where
$\bm{E}(t)=\bm{E}_0  \cos(\omw t) $
is a classical MW electric field, and 
$\edm$ is the EDM operator, which we write 
$\edm = \edmtr \dyad{\one}{\z} + \edmtr^* \dyad{\z}{\one} \equiv \edmtr\sigma^x$, 
where we assume $\edmtr$ is real and introduce
$\sigma^x = \dyad{\one}{\z} + \dyad{\z}{\one}$.
Then, $\Hmw=\hbar\Omega \cos(\omw t)\sigma^x$ with the Rabi frequency $\Omega = -\edmtr\cdot\bm{E}_0/\hbar$.
Assuming the detuning, $\Delta(\Bres) = \omol(\Bres) - \omw$, and Rabi frequency satisfy $|\Delta|,\Omega \ll \omw$, we make the rotating wave approximation (RWA), and obtain the Hamiltonian in the rotating frame for a single molecule (see Appendix~\ref{sec:Derivations} for details),
\begin{align}
 \Hmol 
 &= \Eave(B) \Imat_2 +
  \hbar \Delta \sigma^z/2  + \hbar \Omega \sigma^x /2 \:.
 \label{eq:Hrwa}
\end{align}
Its eigenstates acquire the maximum EDM $\edmtr$ on resonance [see Fig.~\ref{fig:CaF-levels}(c)].
Around resonance, the EDM generated has only second-order sensitivity to fluctuations in the control parameters. 
The effective Hamiltonian Eq.~\eqref{eq:Hrwa} is analogous to single-qubit Hamiltonians encountered in other quantum-information platforms such as trapped ions~\cite{Haffner2008} or superconducting circuits~\cite{Devoret2004}.
It allows single-qubit operations to be performed by changing $\Delta$ or $\Omega$,  each of which can be controlled quickly and robustly in the MW regime. In a many-molecule array, single-molecule gates can be achieved e.g.\ by displacing the molecule of interest in a tweezer array~\cite{Ni2018,Kielpinski2002} or, in an optical lattice, by Stark-shifting the target molecule using an addressing beam~\cite{Weitenberg2011} or crossed laser beams~\cite{Wang2015,Wang2016}.

\section{Simple entangling gate}
We consider next the effect of the magnetic and MW fields on two identical molecules separated by a distance vector $\bm{R}$~\footnote{We implicitly assume that both molecules see the same MW field. This is a good approximation for separations
$R=|\bm{R}| \ll 2\pi c/\omol$  ($\approx 3$~cm for $\omol \approx 2\pi\times 10^9$~rad/s), with $c$ the speed of light.}.
The DDI between the two molecules is
\begin{align}
 \Hddi = 
 \frac{1}{4\pi\epsilon_0 R^3}
 \left( \edm_\mathrm{A}\cdot\edm_\mathrm{B} - 3(\edm_\mathrm{A}\cdot \hat{\bm{R}})\otimes(\edm_\mathrm{B}\cdot\hat{\bm{R}}) \right) ,
 \label{eq:Hddi}
\end{align}
where $\edm_{j}$ is the EDM operator of molecule $j\in\{\text{A}, \text{B}\}$, $\hat{\bm{R}}$ is a unit vector in the direction of $\bm{R}$, and $\epsilon_0$ is the vacuum permittivity.
Recalling the expression for $\edm$ in terms of $\sigma^x$,
we have 
$\edm_\mathrm{A}\cdot\edm_\mathrm{B} = \edmtr^2 \sigma^x_\mathrm{A} \otimes \sigma^x_\mathrm{B}$,
where $\sigma^{\alpha}_j$ is the $\alpha=\{x,y,z\}$ Pauli operator in the qubit space of molecule $j$. 
For a magnetic field along the $z$ axis and MWs linearly polarised along $z$, $\edm_\mathrm{A,B}$ is parallel to the $z$ axis.
In this situation, there are three values of the angle, $\theta$, between $\bm{R}$ and $z$ of particular interest:
    (i) $\theta = \pi/2$,
    (ii) $\theta = \arccos(1/\sqrt{3})$,
    and (iii) $\theta = 0$.
In case (i), the dipoles are side-by-side and we have $\Hddi^{\textrm{(i)}} = \Vddi \sigma_\mathrm{A}^x \otimes \sigma_\mathrm{B}^x$, with $\Vddi =  \edmtrabs^2/(4\pi\epsilon_0 R^3)$.
In case (ii), $\Hddi^{\textrm{(ii)}} = 0$ and 
the coupling vanishes.
Finally, in case (iii), which we use for our numerical simulations, the dipoles are head-to-tail and $\Hddi^{\textrm{(iii)}} = -2\Vddi \sigma_\mathrm{A}^x \otimes \sigma_\mathrm{B}^x$.
For convenience, we write $\Hddi = V  \sigma_\mathrm{A}^x \otimes \sigma_\mathrm{B}^x$, with $V=\eta \Vddi$, with the numerical factor $|\eta| \leq 2$ accounting for the directional dependence.

\begin{figure}[!t] 
  \centering
  \includegraphics[width=\columnwidth]{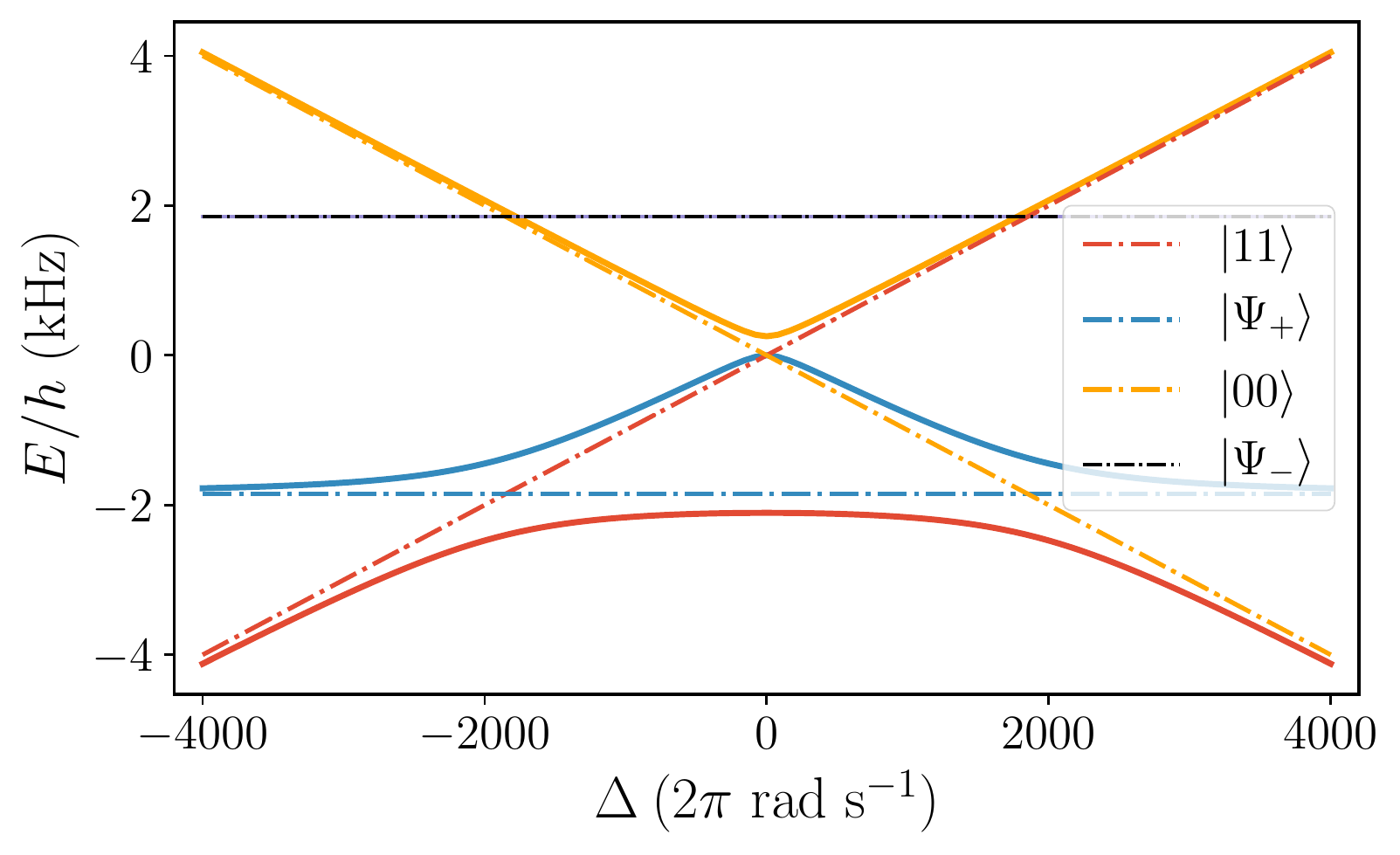}
  \caption{
  Energies of the two-molecule states. 
  Dash-dotted lines represent the eigenenergies of $\Htwomol-2\Eave(B)\Imat_4$ for the same states of CaF as in Fig.~\ref{fig:CaF-levels} for $V=-h\times 1850$~Hz and $\Omega=0$, 
  while solid lines indicate the eigenenergies with the same $V$ and $\Omega = 2\pi\times 731$~rad/s; this DDI strength corresponds to two 1.77~D dipoles 0.8~$\mu$m apart.
  }
  \label{fig:crossing-protocol}
\end{figure}

Assuming now $|\Delta|,\Omega, |V|/\hbar \ll \omw$, 
we make the RWA and find the two-molecule Hamiltonian in the rotating frame
[see Eq.~\eqref{eq:SM-H2mol-Psis}]
\begin{align}
 \Htwomol 
 = \, & 
 2\Eave(B) \Imat_4
 + \hbar\Delta \big( \dyad{\oo}-\dyad{\zz} \big) \nonumber \\
 & + V \big( \dyad{\Psi_+} - \dyad{\Psi_-} \big) \nonumber \\
 & + \bigg[ \frac{\hbar\Omega}{\sqrt{2}} \bigg(
   \dyad{\zz}{\Psi_+} + \dyad{\oo}{\Psi_+} \bigg) + \text{H.c.} \bigg]
 \:.
 \label{eq:H2mol-Psis}
\end{align}
Here 
$\ket{ij}=\ket{i}_\mathrm{A} \ket{j}_\mathrm{B}$ $(i,j \in \{\z,\one\})$,
and 
we introduced the Bell states $\ket{\Psi_{\pm}} = (\ket{\zo} \pm \ket{\oz})/\sqrt{2}$.
It is clear from Eq.~\eqref{eq:H2mol-Psis} that 
$\Htwomol$ does not mix the 
symmetric and antisymmetric subspaces, spanned respectively by $\{ \ket{\oo}, \ket{\Psi_+}, \ket{\zz} \}$ and $\ket{\Psi_-}$.
In the absence of the DDI and MW coupling, the three symmetric states cross at $\Delta=0$.
The DDI shifts $\ket{\Psi_+}$ by $V$, which separates the three-level crossing into three distinct two-level crossings; see Fig.~\ref{fig:crossing-protocol}(a).
These become avoided crossings when $\Omega \ne 0$.
The avoided crossing between $\ket{\zz}$ and $\ket{\oo}$ remains at $\Delta=0$, 
while $\ket{\Psi_+}$ has avoided crossings at $\Delta = \pm V/\hbar$ with $\ket{\zz}$ and $\ket{\oo}$, respectively.

\begin{figure}[t] 
  \includegraphics[width=\columnwidth]{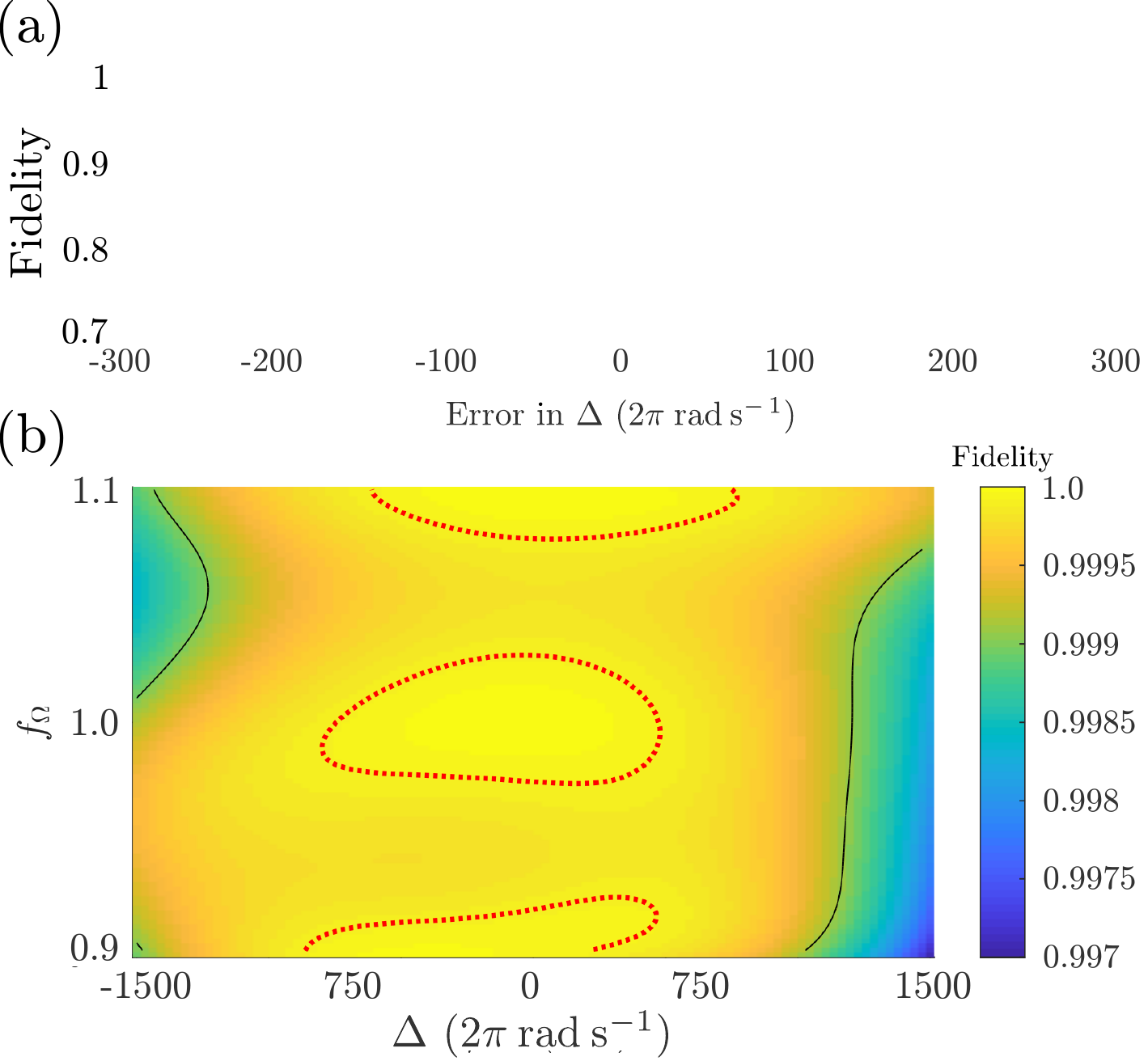}
  \caption{ 
  (a) Fidelity of the protocol with a Gaussian pulse (solid line) when realised with a constant shift in transition frequency $\omol$.
  The horizontal dashed line marks $0.95$ fidelity while vertical dotted lines mark errors of $\pm 100$~Hz.
  Other pulse parameters are: $\tgate = 0.5~\mathrm{ms}$, $\trms = 0.118~\mathrm{ms}$, $\max[\Omega(t)] = 2\pi\times 1200~\mathrm{rad/s}$, and $\Delta = -2\pi \times 1970~\mathrm{rad/s}$.
  (b) Fidelity of the protocol with the GRAPE pulse as a function of
  detuning and relative Rabi frequency $f_\Omega$; see text for details. 
  The red dotted lines indicate $\mathcal{F}=0.9999$ while the black solid lines indicate $\mathcal{F}=0.999$.
  GRAPE optimization parameters are the gate duration $\tgate = 0.5~\mathrm{ms}$
  and $\max[\Omega(t)] = 2\pi\times 50\times10^3$~rad/s.
  }
\label{fig:result-CaF}
\end{figure}

A non-zero DDI thus allows separate addressing of the transitions
$\ket{\oo} \leftrightarrow \ket{\Psi_+}$ and
$\ket{\zz} \leftrightarrow \ket{\Psi_+}$.
The simplest way to show this is to consider a coherent transfer, e.g., from $\ket{\oo}$ to $\ket{\Psi_+}$.
We consider first an implementation using a Gaussian pulse, $\Omega(t)$, of root-mean-squared width $\trms$, at a constant detuning $\Delta$;
numerically, we switch the pulse on and off with a rectangular window function of length $\tgate$.
We require 
a pulse duration $\tgate \gtrsim 2\pi\hbar/|V|$
to be able to resolve the two transitions.
Under these conditions, we achieve a high fidelity for the transfer process, which we define as $\mathcal{F} = \tr( \Ugate^{\dagger} U(\tgate) )$. Here $U(\tgate)$ is
the unitary time-evolution operator on the whole two-qubit space, and 
\begin{equation}
    \Ugate = \begin{pmatrix}
        0 & 1 & 0 & 0 \\
        -1 & 0 & 0 & 0 \\
        0 & 0 & 1 & 0 \\
        0 & 0 & 0 & \exp(i \beta)
    \end{pmatrix}
    \label{eq:Ugate}
\end{equation}
is the desired transformation expressed
in the basis $\{ \ket{\oo}, \ket{\Psi_+}, \ket{\zz}, \ket{\Psi_-} \}$.
The phase $\beta$ is set by the chosen values of $\Delta$ and $\tgate$. 
The main sources of error for this implementation of the entangling gate stem from uncertainties in 
$\omol$ or, equivalently, $\Delta$.
We estimate this by calculating the fidelity of the protocol as a function of a constant error in $\Delta$; see Fig.~\ref{fig:result-CaF}(a). We observe that the fidelity drops to $\approx 0.95$ for detuning errors $\approx 100$~Hz.

\section{Robust entangling gates}
The robustness of the gate can be enhanced by utilizing more general driving schemes that exploit coherences in the full two-qubit Hilbert space~\cite{Glaser2015}.
We use the gradient ascent pulse engineering (GRAPE) algorithm~\cite{Khaneja2005} to design a MW pulse, $\Omega_\textrm{opt} e^{i \xi_\textrm{opt}(t)}$, 
that implements the entangling gate Eq.~\eqref{eq:Ugate}.
This method has the critical advantage of allowing us to obtain pulses that offer robust performance over a range of parameters that span realistic experimental uncertainties.
Specifically, we use GRAPE to obtain step-wise functions $\{ \Omega_\textrm{opt}, \xi_\textrm{opt} \}$ [$\Omega_\textrm{opt}(t)\geq 0$], in $5~\mu$s steps,
that maximize the average fidelity for three values of the Rabi frequency, $\Omega(t) = f_\Omega \Omega_\textrm{opt}(t)$ with $f_\Omega=\{ 0.9, 1.0, 1.1 \}$, and a range of detunings $\leq 1$~kHz; see Appendix~\ref{sec:thermal} for details on our implementation of the GRAPE algorithm.
We show in Fig.~\ref{fig:result-CaF}(b) the fidelity of the time-evolution operator
corresponding to such a GRAPE-optimised pulse, assuming that the molecules are in the motional ground state of their traps. 
The fidelity reaches very high values, $\mathcal{F} > 0.9999$, for 
wide regions of the parameter space, and remains above the quantum-error-correction threshold, $\mathcal{F} > 0.999$, for errors $\lesssim 1$~kHz in detuning and $\leq 10\%$ in the Rabi frequency.

We further extend this approach to deal with a thermal occupation of excited motional states of the trap, if the system is deeply in the Lamb-Dicke regime.
We assume that the system is initially in a product state of internal and motional states, $\rho(t=0) = \rho_\text{int}\otimes \rho_\text{motion}$, and that $\rho_\text{motion}$ is an incoherent superposition of trap states.
We design a pulse that drives the system into 
$\rho(t=\tgate) = (\Ugate \rho_\text{int}) \otimes \rho_\text{motion}$,
and thus implements $\Ugate$ irrespective of motional excitations; see Appendix~\ref{sec:thermal} for details of our modeling of the motional degree of freedom, its coupling with the internal (`qubit') state, and thermal excitations. 
The complexity of the pulse optimisation grows quickly as we require it to generate the same phases for an increasing number of motional states.
The effectiveness of this approach is thus limited to samples cooled to temperatures lower than $h \ftrap / k_\mathrm{B}$, where $\ftrap$ is the trap frequency, so that the population of excited motional states is exponentially suppressed. 
Then, the effect of thermal excitation can be dealt with by truncating the space of motional excitations to a maximum of one in total for the two traps (see Appendix~\ref{sec:pop-motional-states}). 

The fidelity of the pulse, when applied to an initial state with up to one motional excitation, is shown in 
Fig.~\ref{fig:SM-pulse-motion}(b); it is only mildly lower than that in Fig.~\ref{fig:result-CaF}(b) which is applied to the motional ground state. The difference stems mostly from the phase acquired by $\ket{\Psi_-}$, 
which has not been included in the optimisation procedure.
Despite this, the fidelity is still greater than $0.999$ for practically the same region in parameter space.

Scalable application of this protocol within a many-molecule array can be achieved
by spectrally selecting a target pair of nearby molecules.
In an array of tweezers, this can be accomplished, for example, by staggering the intensities of the tweezers to Stark-shift all neighbours with the exception of the chosen pair out of resonance.
In an optical lattice setup, $\omol$ of the target pair can be similarly shifted $>100$~kHz with minimal effect on the confinement using an addressing beam~\cite{Weitenberg2011} or crossed laser beams~\cite{Wang2015,Wang2016}.

\section{Guidelines for state selection}
We expect that the dominant sources of error in implementations of our gates to stem from uncertainties in the transition frequency, $\omol$.
Uncontrolled shifts in $\omol$ arise in experiments due to imperfectly controlled Zeeman and tensor Stark shifts. 
For a magnetic field stability of 1~mG, a 100~Hz stability in $\omol$
requires a transition with magnetic sensitivity below 100~kHz/G.
The transition in CaF highlighted in Fig.~\ref{fig:CaF-levels}(b) has a magnetic sensitivity of only 0.104(4)~kHz/G~\cite{Caldwell2020}, and so is a good choice in this respect. The differential ac Stark shift of states $\ket{\z}$ and $\ket{\one}$ leads to fluctuations in $\omol$ if the intensity of the trap light, $\Itrap$, fluctuates. 
Let $\alpha_{\one,\z}\Itrap$ be the Stark shifts of $\ket{\one}$ and $\ket{\z}$, 
and let $\Delta \alpha = \alpha_{\one}-\alpha_{\z}$ and $\bar{\alpha} = (\alpha_{\one}+\alpha_{\z})/2$. 
We assume $\Delta \alpha \ll \bar{\alpha}$. 
If the intensity changes by $\delta \Itrap$, then $\omol$
changes by $(\Delta\alpha/\bar{\alpha})(\delta \Itrap/\Itrap)U_{\rm trap}/\hbar$, where $U_{\rm trap}$ is the trap depth.
Taking $U_{\rm trap} / h = 1~\mathrm{MHz}$ 
and $\delta \Itrap/\Itrap = 10^{-3}$, a frequency stability of 100 Hz translates to the requirement $(\Delta\alpha/\bar{\alpha}) < 0.1$. Through a careful choice of states, magnetic field magnitude, and polarization angle of the trap light, it is often possible to tune $\Delta \alpha$ to values much smaller than this~\cite{Blackmore2019}.

\section{Discussion and outlook}
A key element of our protocol is the energy shift that the DDI creates in the two-molecule spectrum. This has the same origin as the dipole blockade in Rydberg systems~\cite{Jaksch2000, Lukin2001, Urban2009, Gaetan2009, Johnson2010, Keating2015, Jau2016, Mitra2020}, which is at the core of the Rydberg phase gate~\cite{Jaksch2000}. 
However, our scheme is not susceptible to decoherence and losses in the strongly interacting states because our large-EDM states are low-lying rotational states with negligible spontaneous decay rates ($\lesssim 10^{-8}$~s$^{-1}$~\cite{Ni2018}). This highlights one of the advantages of cold polar molecules for quantum information processing~\cite{Ni2018, Blackmore2019, Sawant2020, Albert2019}.

The idea of switching the DDI that underpins our proposal is similar to the ``dipolar switching'' in Ref.~\cite{Yelin2006}, but our proposal does not involve a static electric field or a third molecular level resonantly coupled with those in the qubit space. As a result, our proposal is simpler to implement and less susceptible to environmental perturbations.

Recently, Ref.~\cite{Ni2018} put forward a proposal for an iSWAP gate between molecules with $\mathcal{F}\geq0.9999$ based on a switchable DDI. This protocol encodes the qubit states in nuclear spin states of the lowest rotational manifold, which are resonantly coupled using MWs to a rotationally excited state with rotation-hyperfine coupling. This allows control of the DDI between two molecules by moving them towards each other and then back apart. Careful timing of this sequence ensures that the two-molecule state acquires the phases required to generate the iSWAP gate.
In addition, all molecular states in~\cite{Ni2018} are insensitive to electric and magnetic fields, providing protection from sources of dephasing.
In contrast to this approach, our proposal involves only two levels and does not call for physical displacements of the molecules; this is simpler and reduces the risk of motional excitation during the gate. 
We take advantage of the large Zeeman splitting between states and the high controllability and stability of modern MW sources to obtain gate times and high fidelities similar to those of Ref.~\cite{Ni2018}.
Earlier, Ref.~\cite{Pellegrini2011} employed optimal control theory (OCT) to design a 
controlled-\textsc{not} (\textsc{cnot}) gate between two polar molecules, achieving a $99\%$ fidelity under ideal conditions; however, the decay of this fidelity against experimental imperfections was not analysed. 
By contrast, the robustness of our scheme to uncertainties in $\omol$, $\Omega$, and to thermal excitations paves the way for practical near-term quantum information processing with polar molecules exploiting their DDI.
Similar ideas of pulse shaping have proven instrumental in state-of-the-art multiqubit gates in 
a variety of experimental platforms~\cite{Schafer2018, Omran2019, Barends2014, Li2019}.

In summary, we have designed a protocol that uses a time-varying microwave field to entangle two polar molecules by controlling the intermolecular DDI. 
Our calculations, based on levels in CaF and RbCs that are precisely known from molecular spectroscopy, demonstrate the possibility of producing maximally entangled two-molecule states with $\geq 99.9\%$ fidelity in less than 1~ms, in a manner that is robust with respect to the main experimental imperfections.
Together with single-molecule gates that can be realised in tweezer arrays or optical lattices by Stark shifting the levels of the target molecule, these results establish the feasibility of building a fault-tolerant quantum processor with cold polar molecules in a scalable optical setup.

Our tools for controlling the states of single molecules and molecular pairs may be applied to advance other quantum technologies with polar molecules.
For example, the possibility of controlling molecular EDMs with easily accessible magnetic and MW fields will expand the range of models that can be simulated using ultracold molecules~\cite{Micheli2006, Gorshkov2011, Gorshkov2013, Covey2018, Blackmore2019, Rosson2020}.
In addition, shaped MW pulses will allow fast control of state-dependent interactions between molecules.
This can be used to explore open questions about the out-of-equilibrium dynamics of power-law-interacting quantum systems, e.g., on quantum thermalisation~\cite{Choi2019} and its interplay with conservation laws~\cite{Neyenhuis2017, Mur-Petit2018}, the transport of excitations~\cite{Nandkishore2017, Deng2018}, or the spreading of correlations~\cite{Foss-Feig2015, Sweke2019}. 
Finally, the large EDMs achievable with MW-dressed molecular eigenstates makes them highly sensitive to external electric fields, which can be exploited to design sensitive detectors of low-frequency ac fields with molecular gases or even single molecules~\cite{Alyabyshev2012, Mur-Petit2015}. 

We acknowledge useful discussions with T.\ Karman, C.\ R.\ Le Sueur, and C.\ S\'anchez-Mu\~noz.
This work was supported by U.K. Engineering and Physical Sciences Research Council (EPSRC) Grants
No.~EP/P01058X/1, 
No.~EP/P009565/1, 
No.~EP/P008275/1, 
and No.~EP/M027716/1, 
and by the European Research Council (ERC) Synergy Grant Agreement No.\ 319286 Q-MAC.
G.B.\ is supported by a Felix Scholarship.

\appendix

\section{Single-molecule and two-molecule Hamiltonians in the rotating wave approximation\label{sec:Derivations}}

We derive here in detail the effective two-molecule Hamiltonian in the rotating wave approximation, in the presence of a bias magnetic field and a nearly resonant microwave field.

\subsection{Single molecule under MW}

We start our discussion from the single-molecule case in the presence of the bias magnetic field, which reduces the effective Hilbert space to that of a two-level system, spanned by Zeeman states that we label $\ket{\one}$ and $\ket{\z}$. As described in the main text, the effective Hamiltonian in the electric dipole approximation is
\begin{align}
 \Hmol^{(1)} 
 =&
 \frac{E_1(B)+E_0(B)}{2}\Imat_2
 + \frac{\hbar\omol(B)}{2} \sigma^z \nonumber \\
 +& \frac{1}{2} [ \hbar \Omega \exp(i \omw t) \sigma^x
    + \mathrm{H.c.} ], 
 \label{eq:SM-Hmol}
\end{align}
where $\Imat_n$ is the $n\times n$ identity matrix, $\sigma^z = \dyad{\one}{\one} - \dyad{\z}{\z}$, $\sigma^x = \dyad{\one}{\z} + \dyad{\z}{\one}$, $E_j(B)$ is the eigenenergy of state $j=0,1$ as a function of magnetic field, and $\omol(B)=(E_1(B)-E_0(B))/\hbar$.

It is now useful to move to the interaction picture with respect to the effective molecular Hamiltonian Eq.~\eqref{eq:SM-Hmol}. To this end, we introduce the unitary operator
$U= \exp(i \omw t/2)\dyad{\one}{\one} + \exp(-i\omw t/2)\dyad{\z}{\z}$
(where we used the orthogonality of $\ket{\z},\ket{\one}$).

When we move to the interaction frame by the transformation $U$, the time evolution of a generic state vector in this frame, $\ket{\psi'} = U \ket{\psi}$ is
\begin{align}
 i\hbar \partial_t \ket{\psi'}
 &= i\hbar\partial_t(U\ket{\psi})
  = i\hbar (\partial_t U) \ket{\psi} + U i\hbar\partial_t\ket{\psi} \nonumber \\
 &= (i\hbar (\partial_t U) U^\dagger + U \Hmol^{(1)} U^\dagger) \ket{\psi'}
 \equiv H^{(I)} \ket{\psi'} ,
 \label{eq:SM-rotFrame}
\end{align}
where we introduce the Hamiltonian in the interaction frame, $H^{(I)}$.
We now introduce $\Eave = (E_1(B) + E_0(B))/2$ and $\Delta(B)=\omol(B)-\omw$.
Under the conditions that $|\Delta|,\Omega\ll 2\omw$, the terms containing the exponentials $e^{\pm 2 i\omw t}$
oscillate very quickly and average to zero on the timescales set by $\Omega^{-1},\Delta^{-1}$, and can therefore be neglected if we are interested in the dynamics only on such timescales; this is the \textit{rotating wave approximation} (RWA). 
Collecting all the terms, the resulting time-independent single-molecule
Hamiltonian in the interaction picture is that in Eq.~\eqref{eq:Hrwa} in the main text, namely
\begin{align}
 \Hmol 
  &= \Eave \Imat_2 +
  \hbar \Delta\sigma^z/2  + \hbar \Omega \sigma^x/2
  \nonumber \\
  &= \begin{pmatrix}
   \Eave + \hbar\Delta/2 & \hbar\Omega/2 \\
   \hbar\Omega/2 & \Eave - \hbar\Delta/2
 \end{pmatrix}
 \label{eq:SM-Hrwa}
\end{align}
in the basis $\{ \ket{\one}, \ket{\z} \}$ where $\ket{\z}$ is the state $\ket{\z}$ shifted up in energy by $\hbar\omw$. 
Its eigenenergies are
\begin{align}
 E_{\stateup, \statedn}
 = \Eave \pm \frac{1}{2} \hbar \sqrt{\Delta^2 + \Omega^2} \:.
 \label{eq:SM-eigenenergies}
\end{align}

\subsection{Two molecules}

We now consider the case of two identical molecules separated by a distance vector $\bm{R}$ and subject to the same magnetic and MW fields. We assume that both molecules see the same MW field, $\bm{E}(t)$, which is a good approximation for separations $R=|\bm{R}| \ll 2\pi c/\omol$ ($\approx 3$~cm for $\omol \approx 2\pi \times 10^9$~rad/s), with $c$ the speed of light.
Therefore, the Hamiltonian describing the two-molecule system is the sum of the two single-molecule Hamiltonians and the dipole--dipole interaction between the two molecules:
\begin{align}
 H_{\text{2mol}}^{(1)} = 
 H_\mathrm{A}\otimes \Imat_2 + \Imat_2 \otimes H_\mathrm{B} + \Hddi \:.
 \label{eq:SM-H2mol}
\end{align}
Here, A, B label the two molecules, and $H_j=\Hmol^{(1)}$ is the Hamiltonian describing the internal space of molecule  $j=\{ \mathrm{A}, \mathrm{B} \}$ in the presence of the magnetic and MW fields [Eq.~\eqref{eq:SM-Hmol}].
As described in the main text, $\Hddi$ can be written
\begin{align}
 \Hddi
 = V \sigma_\mathrm{A}^x \otimes \sigma_\mathrm{B}^x
 = V
 \begin{pmatrix}
   0 & 0 & 0 & 1\\ 
   0 & 0 & 1 & 0\\ 
   0 & 1 & 0 & 0\\ 
   1 & 0 & 0 & 0
 \end{pmatrix}, 
 \label{eq:SM-dAdB01}
\end{align}
where the matrix representation is 
in the basis $\{ \ket{11}, \ket{10}, \ket{01}, \ket{00} \}$ of the two-molecule space; here the two-molecule basis states are defined as product states, $\ket{ i_\mathrm{A} i_\mathrm{B} } = \ket{ i_\mathrm{A} }\otimes\ket{ i_\mathrm{B} } $. 

As with the single-molecule problem, it is useful now to move to the frame rotating at frequency $\omw$, using the unitary transformation 
\begin{align}
 U
 & =\exp(i \gamma) \dyad{11}{11} + \dyad{10}{10} \nonumber \\
 & + \dyad{01}{01} + \exp(-i \gamma) \dyad{00}{00}, 
\end{align}
with $\gamma=\omw t$. 

The non-interacting part of the Hamiltonian transforms to
\begin{align}
 H_{\text{2mol-n.i.}}^{(I)}
 &= 
 i\hbar(\partial_t U)U^\dagger + U(H_\mathrm{A}\otimes \Imat_\mathrm{B} + \Imat_\mathrm{A} \otimes H_\mathrm{B})U^\dagger
 \nonumber \\
 &= 2\Eave \Imat_4 +
 \hbar\Delta ( \dyad{11}{11} -\dyad{00}{00} )\nonumber \\
 & + \large\{ \frac{\hbar\Omega_+}{2}
     \large(\dyad{11}{10} +\dyad{11}{01}+\dyad{10}{00} \nonumber\\
 & \qquad +\dyad{01}{00} \large) + \text{H.c.} \large\} \nonumber\\
 &= \begin{pmatrix}
  2\Eave + \hbar\Delta & \hbar\Omega_+ & \hbar\Omega_+ & 0 \\ 
  \hbar\Omega_- & 2\Eave & 0 & \hbar\Omega_+ \\ 
  \hbar\Omega_- & 0 & 2\Eave & \hbar\Omega_+ \\ 
  0 & \hbar\Omega_- & \hbar\Omega_- & 2\Eave - \hbar\Delta
 \end{pmatrix} , 
\end{align}
with the matrix expression evaluated in the basis $\{ \ket{\oo}, \ket{\oz}, \ket{\zo}, \ket{\zz} \}$.
Here, $\Omega_{\pm}=\frac{1}{2}(1+e^{\pm2i\gamma})\Omega$ and the detuning is  $\Delta(B)=\omol(B)-\omw$ as before.

For the DDI contribution, 
\begin{align}
 U \Hddi U^\dagger
 &= V \{ e^{2i\gamma} \dyad{\oo}{\zz} + \dyad{\oz}{\zo} + H.c. \} \nonumber \\ 
 &= V
 \begin{pmatrix}
   0 & 0 & 0 & e^{2i\gamma}\\ 
   0 & 0 & 1 & 0\\ 
   0 & 1 & 0 & 0\\ 
   e^{-2i\gamma} & 0 & 0 & 0
 \end{pmatrix} , 
\end{align}
with $V=\Vddi$ in case (i), $V=0$ in case (ii), and $V=-2\Vddi$ in case (iii), depending on the orientations of the molecules, as described in the main text.
Hence, collecting all terms, 
\begin{align}
 H_{\text{2mol}}^{(I)}
 &=
 i(\partial_t U)U^\dagger + U H_{\text{2mol}} U^\dagger
 \nonumber \\
 &= 2\Eave \Imat_4 + \hbar\Delta \{ \dyad{11}{11} -\dyad{00}{00} \} 
  \nonumber \\
 & + \Large\{ \frac{\hbar\Omega_+}{2}
     \large(\dyad{11}{10} +\dyad{11}{01}+\dyad{10}{00} \nonumber\\
 & \qquad +\dyad{01}{00} \large) + \textrm{H.c.} \Large\} \nonumber\\
 & \quad + V \{ e^{2i\gamma} \dyad{11}{00} + \dyad{10}{01} + \textrm{H.c.} \} \nonumber \\
 &= 
 \begin{pmatrix}
   2\Eave + \hbar\Delta & \hbar\Omega_+ & \hbar\Omega_+ & e^{2i\gamma}V \\ 
   \hbar\Omega_- & 2\Eave & V & \hbar\Omega_+ \\ 
   \hbar\Omega_- & V & 2\Eave & \hbar\Omega_+ \\ 
   e^{-2i\gamma}V & \hbar\Omega_- & \hbar\Omega_- & 2\Eave -\hbar\Delta
 \end{pmatrix} , 
\end{align}
where the matrix representation is
in the basis $\{ \ket{\oo}, \ket{\oz}, \ket{\zo}, \ket{\zz} \}$.
As before, we assume $\{|\Delta|,\Omega\} \ll \omw$, and also $|V|/\hbar\ll\omw$. If we are interested in the dynamics at timescales longer than $\{1/\Delta,1/\Omega\}$, we can neglect the terms oscillating at $\pm2\omw$, i.e., set $\Omega_{\pm} \mapsto \Omega/2$ and $\exp(2 i\gamma)\mapsto 0$.
In this RWA, the two-molecule Hamiltonian is
\begin{align}
 \Htwomol
 &= 
 2\Eave \Imat_4
 + \frac{\hbar\Delta}{2} \left( \Imat_2 \otimes \sigma^z_\mathrm{B} + \sigma^z_\mathrm{A} \otimes \Imat_2 \right)
 \nonumber \\
 &
 +\frac{\hbar\Omega}{2} \left( \Imat_2 \otimes \sigma^x_\mathrm{B} + \sigma^x_\mathrm{A} \otimes \Imat_2 \right)
 \nonumber \\
 &+ V \left(\sigma^+_\mathrm{A} \otimes \sigma^-_\mathrm{B} + \sigma^-_\mathrm{A} \otimes \sigma^+_\mathrm{B} \right) .
 \label{eq:SM-H2mol-sigmas}
\end{align}
Here $\sigma^+_j = \ket{\one}_j\bra{\z}$ and  $\sigma^-_j=(\sigma^+_j)^\dagger$ are the raising and lowering operators in the qubit space of molecule $j$. 
The terms in Eq.~\eqref{eq:SM-H2mol-sigmas} involving $\Delta$ and $\Omega$ arise from the single-molecule coupling to the MW field, while the last line describes the DDI in the rotating frame. 
This comprises exchange processes of the form $\ket{\zo}\leftrightarrow \ket{\oz}$. Double-flip processes (i.e., transitions $\ket{\oo}\leftrightarrow \ket{\zz}$) involve the absorption or emission of two MW photons and are neglected in the RWA. 

In the basis $\{ \ket{\oo}, \ket{\oz}, \ket{\zo}, \ket{\zz} \}$, this Hamiltonian can be written as the matrix
\begin{align}
 \Htwomol
 &= 
 \begin{pmatrix}
   2\Eave + \hbar\Delta & \hbar\Omega/2 & \hbar\Omega/2 & 0 \\ 
   \hbar\Omega/2 & 2\Eave & V & \hbar\Omega/2 \\ 
   \hbar\Omega/2 & V & 2\Eave & \hbar\Omega/2 \\ 
   0 & \hbar\Omega/2 & \hbar\Omega/2 & 2\Eave -\hbar\Delta 
 \end{pmatrix}.
 \label{seq:SM-H2mol-matrix-10}
\end{align}
The avoided crossing between two levels of the single-molecule problem now translates into a set of avoided crossings among the four two-molecule states.

Finally, we express $\Htwomol$ in the basis $\{ \ket{\oo}, \ket{\Psi_+}, \ket{\zz}, \ket{\Psi_-} \}$, which shows explicitly how the symmetric and antisymmetric subspaces decouple:
\begin{align}
 \Htwomol
 &= 
 2\Eave \Imat_4
 + \hbar\Delta ( \dyad{\oo}-\dyad{\zz} )
  + V \dyad{\Psi_+} 
 \nonumber \\
 & -V \dyad{\Psi_-} + \frac{\hbar\Omega}{\sqrt{2}} \left(
   \dyad{\zz}{\Psi_+} + \dyad{\oo}{\Psi_+} + \text{H.c.}
 \right)
 \nonumber \\
 &= 
 2\Eave \Imat_4 + 
 \begin{pmatrix}
   \hbar\Delta & \hbar\Omega/\sqrt{2} & 0 & 0 \\
   \hbar\Omega/\sqrt{2} & V &  \hbar\Omega/\sqrt{2} & 0 \\
   0 & \hbar\Omega/\sqrt{2} & -\hbar\Delta & 0 \\
   0 & 0 & 0 & -V
 \end{pmatrix} .
 \label{eq:SM-H2mol-Psis}
\end{align}
This expression agrees with Eq.~\eqref{eq:H2mol-Psis} in the main text.
It makes it clear that the DDI shifts the states $\ket{\Psi_{\pm}}$ away from the crossing that would occur on resonance ($\Delta=0$), resulting in the separation into three distinct crossings within the symmetric subspace in Fig.~\ref{fig:crossing-protocol} in the main text.

\subsection{Limits of validity of our approach\label{sec:validity}}

We consider here two potential sources of error outside the derivation above.

First, we consider the possibility that the driving MW pulse may induce an off-resonant transition to a state outside the qubit space, $\{\ket{0}, \ket{1} \}$.
The excitation probability to such states is approximately equal to $\Omega_\text{off}^2/\Delta_\text{off}^2$, where $\Omega_\text{off}$ and $\Delta_\text{off}$ are the Rabi frequency of, and detuning from, the off-resonant excitation. We use Rabi frequencies not larger than 55~kHz, and the closest state is approximately 10 MHz away. The fraction of off-resonant excitation is thus expected to be below $4\times 10^{-5}$ and we neglect it.

A second potential limitation stems from effects beyond the rotating-wave approximation.
The dominant error from the breakdown of the RWA is the Bloch-Siegert shift, i.e., the ac Stark shift due to the counter-rotating terms that have been dropped above~\cite{AllenEberly}. A key element of our protocol is the energy shift that the DDI creates in the two-molecule spectrum. This has the same origin as the dipole blockade in Rydberg systems~\cite{Jaksch2000, Lukin2001, Urban2009, Gaetan2009, Johnson2010, Keating2015, Jau2016, Mitra2020}, which is at the core of the Rydberg phase gate~\cite{Jaksch2000}. 
This shift is of the order of $\Omega^2/\omw$, which in our system is always much less than 1~Hz, and thus very small in comparison with the Rabi frequencies of tens of kHz.
Moreover, this shift is well within the region of high fidelity $\mathcal{F}\geq99.9\%$ offered by our optimised pulses, and its effect on the overall fidelity of our gate is thus negligible.

\section{Effect of thermal excitations\label{sec:thermal}}

\subsection{Population of motional states\label{sec:pop-motional-states}}

We discuss briefly the effect of thermal excitation, resulting in a distribution of the motional quantum number, $n$, of each molecule in its trapping potential. We assume effective cooling towards the motional ground state, so that $\bar{n} \ll 1$. Then, the effect of thermal excitation can be understood by truncating the space of motional excitations to a maximum of $n=1$ in total for both traps. We therefore consider three motional states $\{ \ket{n_\mathrm{A}=0, n_\mathrm{B}=0}, \ket{n_\mathrm{A}=0, n_\mathrm{B}=1},  \ket{n_\mathrm{A}=1, n_\mathrm{B}=0} \}$, where $n_j$ is the number of motional excitations of molecule $j$.

The infidelity due to such motionally excited states is reduced under the assumption that these states are not coupled by the MWs to the ground motional state. 
This is a reasonable approximation given the very small momentum recoil associated with the absorption or emission of a MW photon of frequency $\omol$, i.e., the system is in the Lamb-Dicke regime.
As we demonstrate in the following, under these conditions, it is possible to design an optimal pulse that takes an initial state that is a product of internal and thermal motional states,
$\rho(t=0) = \rho_\text{int}\otimes \rho_\text{motion}$, and drives it into 
$\rho(t=\tgate) = (\Ugate \rho_\text{int}) \otimes \rho_\text{motion}$,
and thus implements the desired quantum gate irrespective of motional excitations.

\begin{figure}[tb]
  \includegraphics[width=\columnwidth]{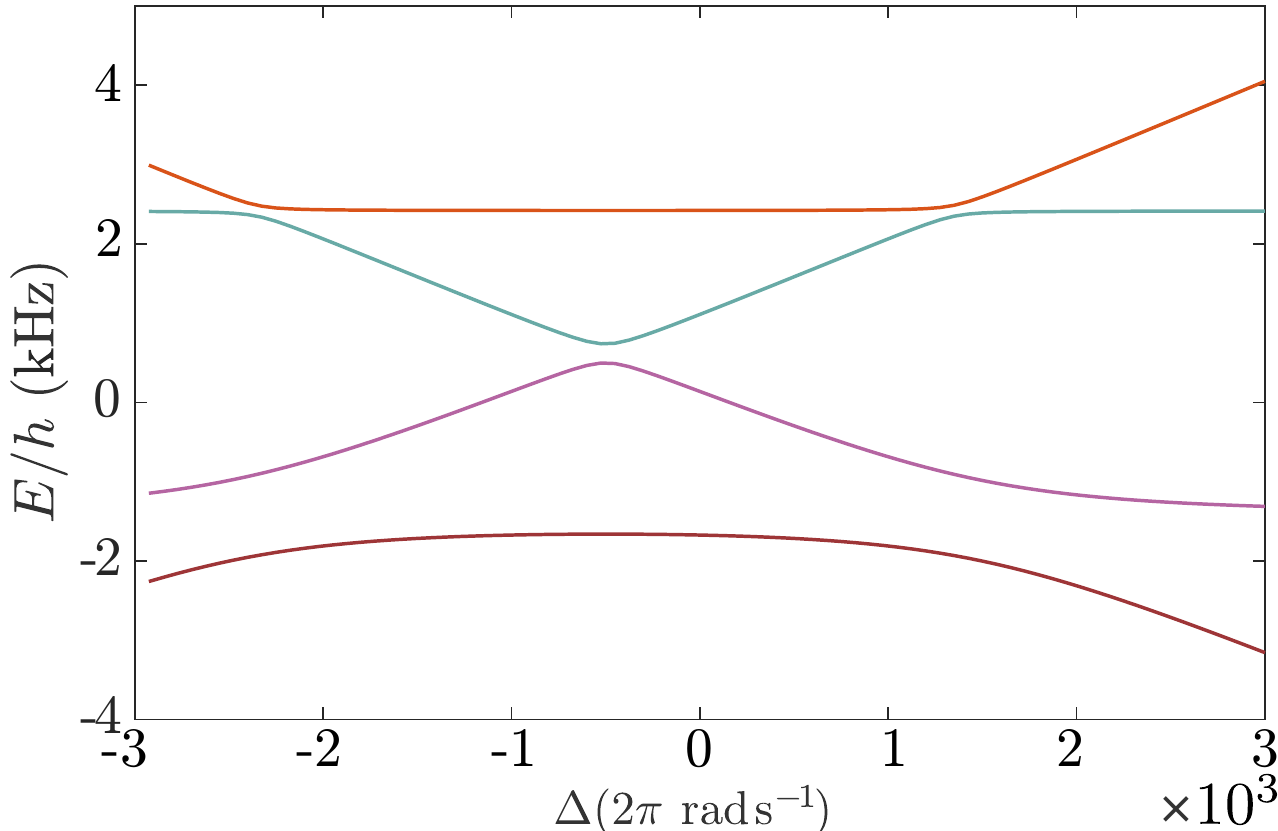}
  \caption{
  Energies of the two-molecule states as a function of $\Delta$, when one molecule is in $n=0$ and the other in $n=1$. A common energy, $2\Eave + h \ftrap$ has been subtracted. The parameters are $V=-h \times 1850$~Hz, $\Omega =  2\pi \times 731$~rad/s, 
  and $\delta \ftrap = 1$~kHz.}
\label{fig:SM-crossings_thermalExcitation}
\end{figure}

To start, let us consider the energies of the two-molecule system as a function of $\Delta$ in the case where one molecule is in the motional ground state, $n_\mathrm{A}=0$, and the other in $n_\mathrm{B}=1$; these are shown in Fig.~\ref{fig:SM-crossings_thermalExcitation}.
Here, we have taken the difference in trap frequency for states $\ket{\z}$ and $\ket{\one}$, $\delta \ftrap$, as 1~kHz.
We choose this large value to illustrate clearly what happens. The other parameters are identical to those used in Fig.~2. There are two main differences compared to Fig.~2(a).
First, the pattern of levels is shifted in $\Delta$ by $\delta \ftrap/2$.
This error in $\Delta$ reduces the fidelity of the entangling protocol by the amount shown in Fig.~3(a) for the simple Gaussian pulse or in Fig.~3(b) for the GRAPE pulse. 
Secondly, the antisymmetric state 
$\ket{\Psi_-}$, which has a constant eigenenergy $V$ when $\Omega=0$ and $\delta \ftrap=0$, no longer completely uncouples from the symmetric subspace.
Instead, avoided crossings open up between $\ket{\Psi_-}$ and $\ket{00}$ at negative $\Delta$, and between  $\ket{\Psi_-}$ and $\ket{11}$ at positive $\Delta$.
They arise because the states $\ket{0,n=0}_{\rm A}\ket{1,n=1}_{\rm B}$ and $\ket{1,n=0}_{\rm A}\ket{0,n=1}_{\rm B}$ are not degenerate when $\delta \ftrap \ne 0$. As a result, terms that can couple $\ket{\Psi_-}$ to the other states no longer cancel in second-order perturbation theory. The widths of these avoided crossings scale with $\Omega$ and with $\delta \ftrap$.
For all relevant values of $\delta \ftrap$, these new avoided crossings are smaller than the one at $\Delta=0$. 
A similar level scheme exists for the states with motional excitations $n_\mathrm{A}=1$, $n_\mathrm{B}=0$ and, higher in energy, for states with $n_\mathrm{A}=n_\mathrm{B}=1$, and so on.

\subsection{Spatial dependence of dipole-dipole interaction\label{sec:spatial-ddi}}

The spatial extent of the molecule wavefunctions in their traps affects the strength of the DDI. To model this, we consider the effect of motion along the line joining the two molecules, i.e., in the direction of $\bm{R}$. The distance operator between the two molecules is given by $\hat{R} = R_\mathrm{e} + \hat{x}_\mathrm{B} - \hat{x}_\mathrm{A}$, where $R_\mathrm{e}$ is the distance between the equilibrium position of the traps and $\hat{x}_j$ is the displacement operator of molecule $j$ from the equilibrium position of its trap along the direction of $\bm{R}$.

In order to calculate how the DDI acts on the internal and motional states, we express the wavefunctions of two given internal $\otimes$ motional states as a function of $x_\mathrm{A}$ and $x_\mathrm{B}$ using the eigenstates of the quantum harmonic oscillator, noting that $\delta \ftrap$ causes the wavefunction of excited motional states to depend on the internal state of the molecules. We also express the DDI operator between internal states $\ket{01}$ and $\ket{10}$ as a diagonal operator in the basis of displacements $x_\mathrm{A}$ and $x_\mathrm{B}$ using $ \Hddi = 
 \frac{-2d^2_{10}}{4\pi\epsilon_0 \hat{R}^3}\ket{10}\bra{01} + H.c.$ for two dipoles aligned head-to-tail. We then use numerical integration over the displacements $x_\mathrm{A}$ and $x_\mathrm{B}$ to find the matrix element of the DDI between the given internal $\otimes$ motional states. After repeating the procedure for all pairs of internal $\otimes$ motional states, these matrix elements were used to build the Hamiltonian in the $12\times12$ basis of four internal states $\otimes$ three motional states. Additional optimisation could similarly consider the motional degrees of freedom perpendicular to $\bm{R}$, but this is beyond the scope of this work.

\begin{figure}[tb]
  \includegraphics[width=\columnwidth]{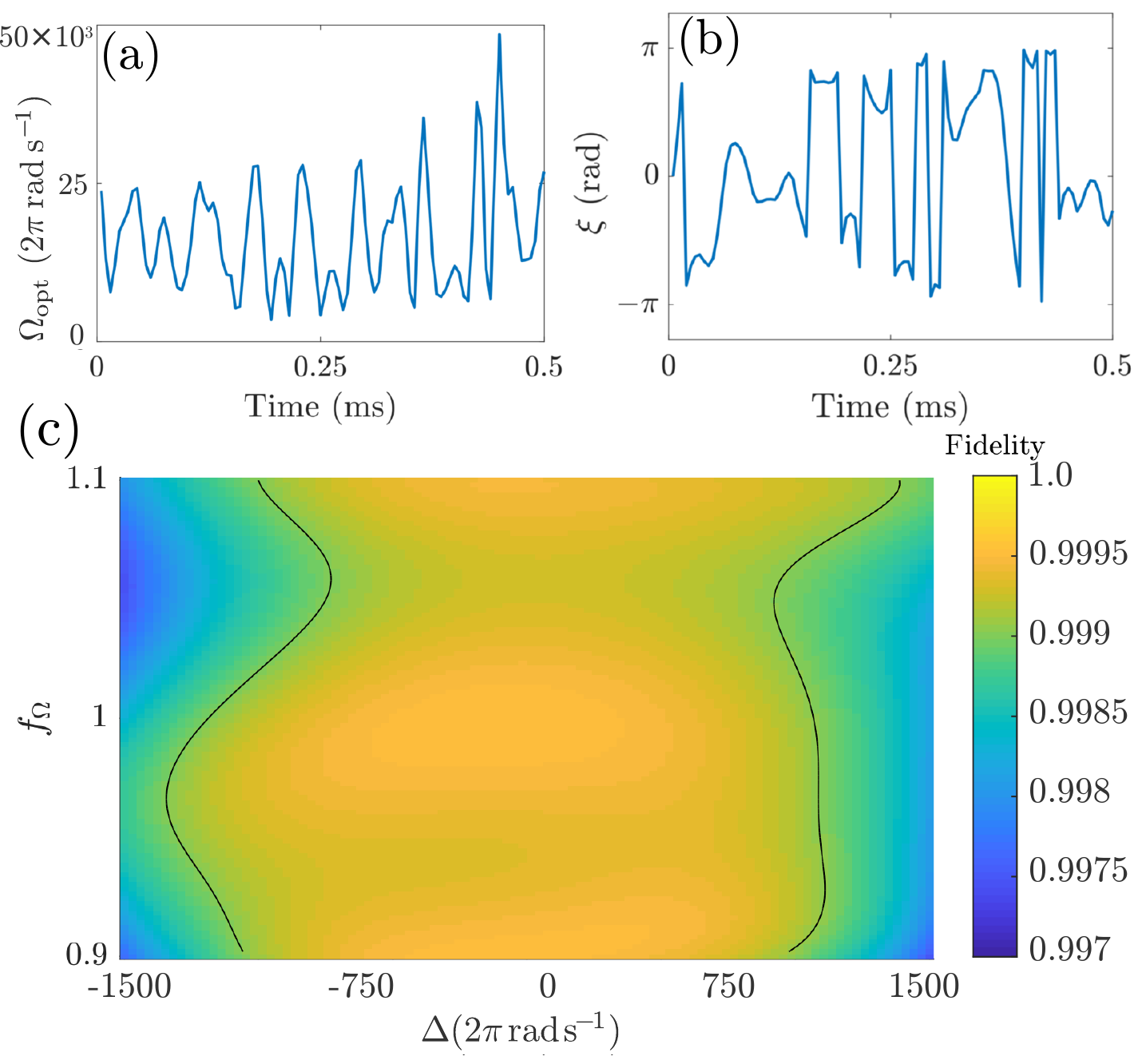}
  \caption{
      Rabi frequency amplitude (a) and phase (b) of the  GRAPE-optimised pulse implementing the entangling gate Eq.~\eqref{eq:SM-U-pulse-thermal}.
      (c) Fidelity as a function of detuning, $\Delta$, and Rabi amplitude error, $f_{\Omega}$.
      The black solid lines indicate $\mathcal{F}=0.999$. Parameters used: $\ftrap = 200$~kHz for one molecule and $204$~kHz for the other. $\delta \ftrap = 500$~Hz for both molecules.
      }
\label{fig:SM-pulse-motion}
\end{figure}

The spatial extent of the harmonic wavefunctions has two effects on the DDI. The first is to modify the expectation value of $\hat{V}$ for a given motional state compared to its value if the dipoles were point particles separated by $R_\mathrm{e}$.
In our calculations, we have $V = -h \times 1847$~Hz for point CaF dipoles separated by $R_\mathrm{e}=0.8~\mu$m, while for the trap parameters used in Fig.~\ref{fig:SM-pulse-motion}, the expectation values $\langle \hat{V} \rangle$ are (rounded to the nearest $h \times 1$~Hz) $-h \times 1862$~Hz in motional state $\ket{n_\mathrm{A}=0, n_\mathrm{B}=0}$ and $-h \times 1877$~Hz in motional states $\ket{n_\mathrm{A}=0, n_\mathrm{B}=1}$ and $\ket{n_\mathrm{A}=1, n_\mathrm{B}=0}$.

The second effect is to couple different motional states. The off-diagonal coupling between the ground motional state $\ket{n_\mathrm{A}=0, n_\mathrm{B}=0}$ and either excited motional state ($\ket{n_\mathrm{A}=0, n_\mathrm{B}=1}$ or $\ket{n_\mathrm{A}=1, n_\mathrm{B}=0}$) is $h \times 145$~Hz.
This is much smaller than the energy difference between the ground and excited motional states, $h \times \ftrap$ ($\ftrap \approx 200$~kHz). 
It follows that population transfer between ground and excited motional states induced by the DDI is of the order of $|145/(200\times10^3)|^2 \sim 10^{-6}$.

Similarly, there is a weak coupling of $h \times 15$~Hz between the excited states $\ket{n_\mathrm{A}=0, n_\mathrm{B}=1}$ and $\ket{n_\mathrm{A}=1, n_\mathrm{B}=0}$. 
By the same reasoning as in the previous paragraph, this coupling leads to a population transfer of order $[15/(\ftrap^\mathrm{A}-\ftrap^\mathrm{B})]^2$, where $\ftrap^j$ is the trapping frequency of molecule $j\in\{\rm A,B\}$.
It follows that a small difference in trapping frequencies $|\ftrap^\mathrm{A}-\ftrap^\mathrm{B}|\gtrsim 2$~kHz is sufficient to bound the population transfer between $\ket{n_\mathrm{A}=0, n_\mathrm{B}=1}$ and $\ket{n_\mathrm{A}=1, n_\mathrm{B}=0}$
to $\mathcal{O}(10^{-4})$.

We have included both these shifts and couplings between motional states in our numerical calculation of the propagator $U(t)$ and hence the gate fidelity $\mathcal{F}$.
The results are shown in Fig.~\ref{fig:SM-pulse-motion}(c).
These results should be contrasted with the results for the fidelity in Fig.~\ref{fig:result-CaF}(b) of the main text, which shows the fidelity corresponding to the same GRAPE-optimised pulse when considering only the ground motional space, $\ket{n_\text{A}=0, n_\text{B}=0}$.
Comparison of these calculations indicates that fidelity $\mathcal{F} \geq 0.999$ is still accessible in a wide region of parameter space when the initial state contains a small contribution from motionally excited states.

\subsection{Internal-state and motional-state separation\label{sec:internal-motional-separation}}

If the two-molecule state is a product of internal and motional states, the unitary operator
\begin{align}
 \Udesired = \Ugate \otimes \Imat_3
 \label{eq:SM-U-pulse-motion}
\end{align}
generates the desired gate in the internal space for all three motional
subspaces for generic states that are a product of internal and motional states.
In practice, it is difficult to design a pulse that ensures this equality in phases, but we can still design a pulse assuming the relevant situation that the motional part is an incoherent superposition of motional states, such as a thermal state. Then, we can write the two-molecule state in the form
\begin{align}
    \rho &= \rho_{\rm int} \otimes \rho_{\rm motion},
\end{align}
with
$ \rho_{\rm motion} = \sum_{n_\mathrm{A}, n_\mathrm{B}} p_{n_\mathrm{A}, n_\mathrm{B}} \dyad{(n_\mathrm{A}, n_\mathrm{B})}$.
Consider now a unitary operator of the form
\begin{align}
 U &= \Ugate \otimes \Umot,
 \label{eq:SM-U-pulse-thermal}
\end{align}
with $\Umot = \mathrm{diag} [ \exp( i \phi_{n_\mathrm{A}n_\mathrm{B}} ) ]$ a diagonal matrix.
The unitary operator $U$ generates a different phase in each motional subspace, but all internal states acquire the same motional phase within that motional subspace $\ket{(n_\mathrm{A},n_\mathrm{B})}$, i.e., 
apart from the motional phases, all internal states are transformed according to the desired $\Ugate$.
The action of $U$ in Eq.~\eqref{eq:SM-U-pulse-thermal} on $\rho$ is
\begin{align}
U^\dagger \rho U
 & = \left( \Ugate^\dagger \rho_\mathrm{int} \Ugate \right) \otimes 
 \left( \Umot^\dagger \rho_\mathrm{motion} \Umot \right) \nonumber \\
 =&\, \Ugate^\dagger \rho_\mathrm{int} \Ugate \otimes \nonumber \\
 & \!\!\! \sum_{n_\mathrm{A},n_\mathrm{B}} p_{n_\mathrm{A} n_\mathrm{B}} 
    e^{-i \phi_{n_\mathrm{A}n_\mathrm{B}}} \dyad{(n_\mathrm{A},n_\mathrm{B})} e^{i \phi_{n_\mathrm{A}n_\mathrm{B}}} \nonumber \\
 =&\, \Ugate^\dagger \rho_\mathrm{int} \Ugate \otimes \rho_\mathrm{motion} \nonumber \\
 \equiv&\, \Udesired^\dagger \left( \rho_\mathrm{int} \otimes \rho_\mathrm{motion} \right) \Udesired.
\end{align}
This means that, as long as the motional part is an incoherent superposition of motional eigenstates, it suffices to design a pulse that implements our target gate $\Ugate$ with high fidelity in each motional subspace separately, as the motional phases $\phi_{n_\mathrm{A}, n_\mathrm{B}}$ will not appear in the transformed state, $U^\dagger \rho U$.

We show in Figs.~\ref{fig:SM-pulse-motion}(a) and \ref{fig:SM-pulse-motion}(b) the Rabi frequency amplitude and phase of a pulse designed in this way. 
Figure~\ref{fig:SM-pulse-motion}(c) shows the fidelity of the time evolution generated with this pulse as a function of detuning and relative Rabi frequency.
We calculate the fidelity by numerically 
determining the motional phases, $\phi_{n_\mathrm{A},n_\mathrm{B}}$, generated
and using $\mathcal{F} = \tr[ (\Ugate \otimes \Umot)^{\dagger} U(\tgate)]$,
with $U(\tgate)$ the unitary operator evolving the two-molecule state in the full $12\times12$ (internal$\,\otimes\,$motional) space. 
We observe that it is possible to achieve fidelities $\mathcal{F} \geq 0.999$, which supports the robustness of our approach to entangle two molecules even in the presence of some residual incoherent motional excitation.

In practice, the effectiveness of this approach is constrained to well cooled samples, $T \ll h \ftrap / k_\textrm{B}$, because the complexity of the pulse optimisation grows quickly as one requires it to generate the same phases on the internal states for an increasing number of motional state blocks; cf.\ Eq.~\eqref{eq:SM-U-pulse-thermal}.

\section{Summary of GRAPE algorithm implementation\label{sec:grape}}

Gradient ascent pulse engineering (GRAPE)~\cite{Khaneja2005} is a powerful optimal control algorithm used to design control pulses which can generate unitary dynamics in a quantum system. A quantum system interacting with time-dependent electromagnetic fields can be described by the Hamiltonian
\begin{align}
  H = H_0 + H_\mathrm{c}(t) \:.
\end{align}
Here, $H_0$ is the time-independent internal Hamiltonian whereas $H_\mathrm{c}(t)$ is the time-varying external control field. In our system, we employ GRAPE to design a pulse that implements the desired gate $\Ugate$ in the $3\times 3$ symmetric internal space. Afterwards, we assess the fidelity of the gate by evolving the two-molecule state within the whole $4\times 4$ internal space.

In this approach, the forms taken by $H_0$ and $H_\mathrm{c}(t)$ are as follows
\begin{align}
    H_0
    &= \begin{pmatrix}
    \Delta & 0 & 0 \\
    0 & V & 0 \\
    0 & 0 & -\Delta
    \end{pmatrix}
    \\
    H_\mathrm{c}(t)
    &=
    \Omega_x(t) I_x + \Omega_y(t) I_y
\end{align}
in the basis $\{ \ket{\oo}, \ket{\oz}, \ket{\zo} \}$.
Here $\Omega_x(t)$ and $\Omega_y(t)$ are MW frequency control fields along the $X$ and $Y$ quadratures described by Pauli spin-1 operators $I_x$ and $I_y$ respectively. We shall work in natural units where $\hbar=1$. The time evolution of this system is
given by the propagator $U(t)$.
We want to evolve the system in time by tuning the control fields $\Omega_x(t)$ and $\Omega_y(t)$ such that the propagator $U(t)$ is as close as possible to the desired target unitary $U_T$. In other words, we want to maximize the fidelity given by
\begin{align}
    \mathcal{F} = |\braket{U_T}{U(t)}|^2.
\end{align}

The GRAPE algorithm is an efficient numerical algorithm to calculate the control fields $\Omega_x(t)$ and $\Omega_y(t)$ which maximize the fidelity $\mathcal{F}$.
Since the terms in the Hamiltonian $H$ are non-commuting, calculating the propagator $U(t)$ is  difficult. To deal with this, the total evolution time $T$ is discretized into $N$ time steps of duration $dt=T/N$. The heart of the GRAPE algorithm lies in efficiently calculating the gradient of control fields at each time step as described in~\cite{Khaneja2005}. The convergence of the GRAPE algorithm can be accelerated by using the Broyden-Fletcher-Goldfarb-Shanno (BFGS) iterative method, which employs second-order gradients to solve the nonlinear optimization problem underlying the GRAPE algorithm; the combined algorithm is known as BFGS-GRAPE~\cite{DeFouquieres2011, Machnes2011}. 
We use this approach to design the MW pulses. 

To deal with potential variations or uncertainties in the level splitting, $\Delta$, as well as in the control fields, $\Omega_x(t)$ and $\Omega_y(t)$,
we require the output control fields to maximize the fidelity over a range of $\Delta$ and $\Omega(t) = \sqrt{ \Omega_x^2(t) + \Omega_y^2(t)}$ using averaging techniques as described
in~\cite{Khaneja2005}.
The optimal Rabi frequency displayed in Figs.~\ref{fig:SM-pulse-motion}(a) and \ref{fig:SM-pulse-motion}(b) (parametrized as $\Omega_{\text{opt}}(t)=\sqrt{ \Omega_x^2(t) + \Omega_y^2(t) }$
and
$\xi(t)=\arctan[ \Omega_y(t) / \Omega_x(t) ]$)
is thus the Rabi frequency that maximizes the average fidelity for 0.9, 1.0 and 1.1 times the nominal MW Rabi frequency.

Let us emphasize again that, while the output of the GRAPE optimization is designed taking into account the $3\times3$ symmetric space, the fidelities reported in Fig.~\ref{fig:result-CaF} of the main text have been calculated  evolving the two-molecule state within the whole $4\times 4$ space.

In the preceding discussion we have described the algorithm to obtain the optimal control fields taking into account only the internal dynamics of the two-molecule system.
As discussed in Appendix~\ref{sec:internal-motional-separation}, at sufficiently low excitation energies in which no more than one motional excitation is present in the system, the DDI leads to small shifts in the energy of the states. Importantly, it also leads to weak couplings between motional states. 
We used the BFGS-GRAPE algorithm to design a pulse that maximises the average fidelity $\mathcal{F}$ within the internal space
in the three separate motional spaces, $\{ \ket{n_\mathrm{A}=0, n_\mathrm{B}=0}, \ket{n_\mathrm{A}=0, n_\mathrm{B}=1},  \ket{n_\mathrm{A}=1, n_\mathrm{B}=0} \}$, taking into account the slightly different internal-space level splittings induced by the DDI.
The optimised pulse was then used to calculate the time evolution with the full Hamiltonian that includes both the DDI shifts and coupling between motional states. That is, the numerically calculated time-evolution propagator $U(t)$ includes processes like
\begin{align*}
 \ket{01}_\text{internal}
 & \otimes\ket{n_\mathrm{A}=1, n_\mathrm{B}=0}_\text{motion}
 \to \nonumber \\
 & \to 
 \ket{10}_\text{internal}\otimes\ket{n_\mathrm{A}=0, n_\mathrm{B}=1}_\text{motion}
 \:,
\end{align*}
that can be understood as ``phonon-induced spin flips.''
The numerical results for the process fidelity shown in Fig.~\ref{fig:SM-pulse-motion} demonstrate that the pulse optimised in this way is robust with respect to such processes, as long as the low-excitation requirement is fulfilled and there is a sufficient difference in the trap frequency of the molecules to make these processes off-resonant as described in Appendix~\ref{sec:spatial-ddi}.


\section{Entangling gate calculations for $^{87}$R\lowercase{b}$^{133}$C\lowercase{s} molecules\label{sec:RbCs}}

The same coupling scheme and entangling gate can also be applied to RbCs. For this molecule, we label the states $\ket{(N, m_F )_j}$ , where $j$ indexes levels with the same $N$ and $m_F$ in ascending order of energy, starting from $j=0$. We set a magnetic field of $181.5$~G to separate the Zeeman states and choose $\ket{\z}=\ket{(0,4)_1}$ and $\ket{\one}=\ket{(1,4)_1}$ as our qubit states. These levels have a transition dipole moment $\edmtrabs=0.482$~D when $\pi$ polarised microwaves of angular frequency
$\omol = 2\pi \times 980.138\times10^6$~rad/s 
are applied in a tweezer trap of intensity $5$~kW/cm$^{2}$.
As for the CaF states discussed in the main text, these states are chosen to optimise the stability of $\omol$ to fluctuations in the tweezer light intensity and magnetic fields.

The calculations for the two-qubit gate in the absence of motional excitations depend in practice only on the magnitude of $\Vddi$.
It follows that the same optimised pulse used for CaF can be used with RbCs, once the Rabi field amplitude and times are scaled accordingly:
\begin{align}
  \Omega^\mathrm{RbCs}(t)
  &=
  \Omega^\mathrm{CaF}
    \left( t \zeta \right) / \zeta, \\
  \zeta
  &=
  \left|\frac{\Vddi^\mathrm{RbCs}}{\Vddi^\mathrm{CaF}}\right|
  =
  \left(\frac{\edmtrabs^\mathrm{RbCs}}{\edmtrabs^\mathrm{CaF}} \right)^2
  \left(\frac{R^\mathrm{CaF}}{R^\mathrm{RbCs}} \right)^3.
\end{align}
Thus, we can generate the same entangling gate between two RbCs molecules in a time $\tgate^\mathrm{RbCs} = \tgate^\mathrm{CaF}/\zeta$
using Rabi frequencies scaled by a factor $\zeta$.
The fidelity shows the same robustness against detuning and noise in the Rabi frequency, $f_\Omega$, behaviour as in Fig.~\ref{fig:result-CaF} in the main text, apart from a rescaling of the detuning axis.
For example, given $\edmtrabs=0.482$~D, if the two RbCs molecules are trapped in optical tweezers $0.8~\mu$m apart (as for CaF in the main text), we obtain
$\zeta = ( 0.482 / 1.77 )^2 \approx 0.074$:
the entangling gate can be generated in $(0.54~\mathrm{ms})/\zeta \approx 7.3$~ms.
If the molecules are trapped instead in an optical lattice with lattice constant 532~nm, 
$\zeta = ( 0.482 / 1.77 )^2 \times (800/532)^3 \approx 0.25$, and the entangling gate can be run in $\approx 2.2$~ms.



\end{document}